\documentclass[twocolumn, prd, superscriptaddress,nofootinbib]{revtex4-1}

\usepackage{array,mathtools,booktabs}
\newcolumntype{C}{>{$}c<{$}}
\AtBeginDocument{
\heavyrulewidth=.08em
\lightrulewidth=.05em
\cmidrulewidth=.03em
\belowrulesep=.65ex
\belowbottomsep=0pt
\aboverulesep=.4ex
\abovetopsep=0pt
\cmidrulesep=\doublerulesep
\cmidrulekern=.5em
\defaultaddspace=.5em
}

 \usepackage{appendix}
\usepackage{hyperref}
\usepackage[utf8]{inputenc}
\usepackage[dvipsnames]{xcolor}
\usepackage{amsmath,amssymb}
\hypersetup{
    colorlinks,
    linkcolor={red!80!black},
    citecolor={green!60!black},
    urlcolor={blue!60!black}
}
\usepackage{graphicx}

\newcommand{\OO}{\mathcal{O}}

\newcommand{\GeV}{\text{GeV}}

\newcommand{\x}[1]{\ensuremath{\text{#1}}} % text mode shortcut

\newcounter{qnumber}

\newcounter{qnumber2}

\newcounter{qnumber3}

\def\r{\right)}
\def\l{\left(}

\begin{document}

%\preprint{APS/123-QED}

\title{Axion production and detection with superconducting RF cavities}
%\title{A Resonant Experiment for Axion enLightenment with a Superconducting OuroBoros (A REAL SOB)}

\author{Ryan Janish}
\affiliation{Berkeley Center for Theoretical Physics, Department of Physics,
University of California, Berkeley, California 94720, USA}

\author{Vijay Narayan}
\affiliation{Berkeley Center for Theoretical Physics, Department of Physics,
University of California, Berkeley, California 94720, USA}

\author{Surjeet Rajendran}
\affiliation{Department of Physics \& Astronomy, The Johns Hopkins University, Baltimore, MD  21218, USA}
\affiliation{Berkeley Center for Theoretical Physics, Department of Physics,
University of California, Berkeley, California 94720, USA}

\author{Paul Riggins}
\affiliation{Berkeley Center for Theoretical Physics, Department of Physics,
University of California, Berkeley, California 94720, USA}
\begin{abstract}

We propose a novel design of a laboratory search for axions based on photon regeneration with superconducting RF cavities.
Our particular setup uses a toroid as a region of confined static magnetic field, while production and detection cavities are positioned in regions of vanishing external field.
This permits cavity operation at quality factors of $Q \sim 10^{10} - 10^{12}$.
The limitations due to fundamental issues such as signal screening and back-reaction are discussed, and the optimal sensitivity is calculated.
This experimental design can potentially probe axion-photon couplings beyond astrophysical limits, comparable and complementary to next generation optical experiments.

\end{abstract}
\maketitle

%%%%%%%%%%%%%%%%%%%%%%%%%%%%%%%%%%%%%%%%%%%%%%%%%%%%%%%%%%%%%%%%
\section{Introduction}
\label{sec:intro}

Axions are well-motivated additions to the standard model (SM).
They provide an elegant solution to the strong CP problem~\cite{Peccei:1977hh, Peccei:1977ur, Weinberg:1977ma, Wilczek:1977pj}, are a natural dark matter candidate~\cite{Preskill:1982cy, Abbott:1982af,Dine:1982ah}, can relax naturalness problems~\cite{Graham:2015cka, Graham:2019bfu}, and appear generically in theories of quantum gravity~\cite{Svrcek:2006yi, Arvanitaki:2009fg}.
Purely laboratory searches for axions are thus an important experimental front.
Given that axions can naturally be very light and have suppressed interactions with the SM, a promising approach is to search for the coherent interaction of a classical axion field with electromagnetic (EM) fields~\cite{Sikivie:1983ip}.

Photon regeneration, or ``Light Shining Through Walls" (LSW), experiments make use of axion-photon oscillations in a transverse magnetic field to convert photons into axions that can traverse an optical barrier and then convert back into detectable photons~\cite{VanBibber:1987rq}.
Small axion-photon conversion probabilities are overcome by the use of resonators to sustain large EM fields~\cite{Sikivie:2007qm}.
This is the basis of experiments such as the Any Light Particle Search (ALPS)~\cite{Ehret:2009sq, Ehret:2010mh, Bahre:2013ywa}, which employ optical cavities aligned with dipole magnets over a long baseline.
LSW can also be done at radio frequencies (RF)~\cite{Hoogeveen:1992nq, Jaeckel:2007ch,Caspers:2009cj}, as in the CERN Resonant Weakly Interacting sub-eV Particle Search (CROWS)~\cite{Betz:2013dza}, by producing and detecting the axion through excited modes in matched RF cavities subject to an external magnetic field.
While interesting, current constraints from LSW experiments are less stringent than those due to stellar cooling or searches for solar axions (see~\cite{Graham:2015ouw} for a review).

We propose a novel design for an axion LSW experiment using high-$Q$ superconducting RF (SRF) cavities, which can in principle reach beyond these astrophysical bounds.
SRF cavities provide an opportunity for a significantly enhanced axion search due to their extremely large quality factors; however, they must be isolated from large magnetic fields in order to avoid catastrophic SRF degradation.
This requires several qualitative modifications from previous setups, most notably the use of a sequestered axion-photon conversion region containing a confined static magnetic field while production and detection cavities are positioned in regions of vanishing static field.
The focus of this paper is to determine the fundamental factors that affect the sensitivity of such an experimental design---a more detailed consideration of experimental strategies is left to future work.
We calculate the optimal signal strength and irreducible noise sources, and find the proposed setup capable of probing axion-photon couplings beyond astrophysical limits and with a reach comparable and complementary to next generation optical experiments.

%%%%%%%%%%%%%%%%%%%%%%%%%%%%%%%%%%%%%%%%%%%%%%%%%%%%%%%%%%%%%%%%
\section{Conceptual Overview}
\label{sec:overview}

LSW searches rely on the axion EM interaction, given by the effective Lagrangian
\begin{align}
\label{eq:lagrangian}
 - \frac{1}{4} F_{\mu \nu}F^{\mu \nu} +
  \frac{1}{2} (\partial_\mu a)^2 -
  \frac{1}{2} {m_a}^2 a^2
  - \frac{1}{4}g a F_{\mu \nu} \widetilde{F}^{\mu \nu},
\end{align}
where $a$ is the axion field of mass $m_a$, $\widetilde{F}^{\mu \nu} = \epsilon^{\mu \nu \rho \sigma}F_{\rho \sigma}$, and $g$ is the axion-photon coupling.
In the limit of classical fields, an axion obeys the equation of motion
 \begin{align}
\label{eq:KG}
 \l \Box +m_a^2 \r a = -g \vec{E} \cdot \vec{B},
\end{align}
and modifies Maxwell's equations:
\begin{align}
\label{eq:ME1}
   \vec{\nabla} \cdot \vec{E} &=  - g \vec{B} \cdot {{\vec{\nabla} a}}, \\
  \label{eq:ME2}
   \vec{\nabla} \times {\vec{B}} &=
   \frac{\partial {\vec{E}}}{\partial t} - {g} \l{\vec{E}} \times \vec{\nabla} a - {\vec{B}} {\frac{\partial a}{\partial t}}\r.
\end{align}
We will generally consider any light, neutral pseudoscalar $a$ and treat $\{m_a , g \}$ as independent parameters.

In an RF LSW experiment such as CROWS~\cite{Betz:2013dza}, a production cavity sources axions through a non-vanishing $\vec{E} \cdot \vec{B}$, where $\vec{E}$ is the electric field of an excited cavity mode and $\vec{B}$ is an external, static magnetic field.
These axions propagate into a detection cavity where, again in the presence of a static magnetic field, they excite an identical frequency mode in the detection cavity.
The signal power that can be extracted is~\cite{Hoogeveen:1992nq}:
\begin{align}
\label{eq:RFsignal}
    P_\text{signal}  = P_\text{input} Q_\text{pc} Q_\text{dc} \l\frac{g B_0}{f}\r^4 |G|^2.
\end{align}
Here $Q_\text{pc}$ and $Q_\text{dc}$ are the loaded quality factors of production and detection cavities, $f \approx \text{GHz}$ is the frequency of the excited modes, $P_\text{input}$ is the driving RF power delivered to the production cavity, and $B_0$ is the static field penetrating both cavities.
$|G|$ is a form factor which depends on the arrangement of the cavities, choice of modes, etc.
This is roughly constant for $m_a  \lesssim  2 \pi f$, and is exponentially suppressed for larger masses.

The quality factors $Q$ of both cavities are key factors in determining the sensitivity of such an experiment.
For normal conducting cavities, $Q \sim 10^5 - 10^6$, however advances in SRF technology have led to the development of superconducting cavities with $Q~\sim 10^{10} - 10^{12}$ which have application in particle accelerators~\cite{Grassellino:2013nza}.
It is worthwhile to consider whether these can be leveraged in an axion LSW search~\cite{srf1,srf2}.%
\footnote{See~\cite{Bogorad:2019pbu} for a proposal to detect axions with SRF cavities that is quite distinct from ours.}
A simple replacement of the RF cavities%
\footnote{It is actually not obvious whether a larger signal is obtained by replacing both cavities or only the detection cavity, see Sec.~\ref{sec:source}---we choose here to study an SRF production cavity as it involves some novel considerations.} %
in the above arrangement with SRF cavities does not work---an external $B_0$ greater than the critical field $\sim 0.2 \, \x{T}$, at which flux penetrates the cavity, would result in excessive dissipation and degrade $Q$.

\begin{figure}
\centering
\includegraphics[width=8cm]{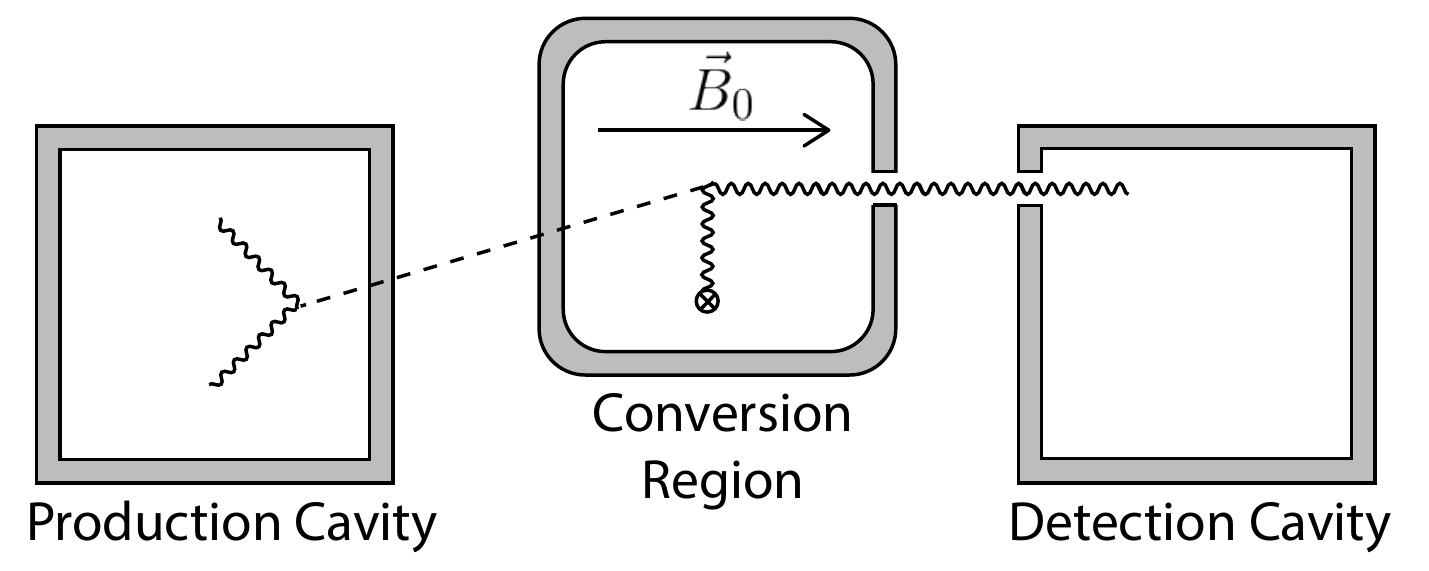}
\caption{Basic elements of an axion LSW experiment using SRF cavities and a conversion region of confined static magnetic field, to be contrasted with an RF cavity experiment such as CROWS~\cite{Betz:2013dza}.}
\label{fig:lsw-schematic}
\end{figure}

This problem is avoided by placing production and detection SRF cavities in regions of vanishing static field while confining a large, static magnetic field in a distinct conversion region, depicted schematically in Fig.~\ref{fig:lsw-schematic}.
The basic elements of an SRF LSW experiment as follows:
\begin{enumerate}
  \item Axions are sourced in a production cavity free of any external field.
  \item The axions then convert into photons in an isolated region of static magnetic field.
  \item The resulting photons propagate out of the conversion region---that is, the axion-induced fields must not also be screened by the conductors which confine the large static field.
  \item Any resulting RF signal is coupled to and amplified by an SRF detection cavity.
  \end{enumerate}

We discuss a possible design that is able to realize all these conditions, and in what follows we will use it to determine the optimal sensitivity of an axion SRF search.

(1) A specific mode or set of modes in the production cavity is driven such that $\vec{E} \cdot \vec{B}$ does not identically vanish.

(2) A static $B_0$ is generated and confined by DC current-carrying superconducting wires wrapped to form a toroidal enclosure.

(3) There is a gap in this enclosure, preventing the toroid from acting as a shielding cavity for the axion-induced fields.
Our use of a gapped toroid is inspired by its related use in experiments (ABRACADABRA and DM Radio) searching for dark matter axions~\cite{Kahn:2016aff, Chaudhuri:2014dla}.

(4) The axion-induced fields are coupled to the detection cavity inductively via an outside pickup loop.
Here, we must properly account for the back-reaction of the amplified signal onto the toroid.
We emphasize that a realistic implementation would require a more detailed signal field read-out mechanism in order to maintain a large effective $Q$ on the detection side.

%%%%%%%%%%%%%%%%%%%%%%%%%%%%%%%%%%%%%%%%%%%%%%%%%%%%%%%%%%%%%%%
\section{Determining the Axion Signal}
\label{sec:signal}

\subsection{SRF axion source}
\label{sec:source}

The axion field produced by an SRF cavity is given by~\eqref{eq:KG}, with the EM fields on the right-hand-side being those of the driven cavity modes.
We focus on one frequency component $\omega$ of this $\vec{E}\cdot\vec{B}$, which may arise from a single cavity mode with frequency $\omega/2$ or from two distinct modes driven together whose frequency sum or difference is $\omega$:
 \begin{align}
 \label{eq:a-field}
     a\l \vec{x}, t \r = -g \, e^{i \omega t} \,
     \int_\x{pc} \, d^3 \vec{y} \, \frac{e^{i k_a |\vec{x} - \vec{y}|}}{4 \pi |\vec{x} - \vec{y}|} \l \vec{E}\cdot\vec{B} \r_{\omega},
 \end{align}
where $k_a = \sqrt{\omega^2 - m_a^2}$ is the axion momentum and the subscript $\omega$ on $\vec{E}\cdot\vec{B}$ indicates restriction to a single frequency component.
The integration $\vec{y}$ is taken over the volume of the production cavity and $\vec{x}$ indicates any point in space, e.g., within the conversion region.
The driven modes must be chosen such that $(\vec{E}\cdot\vec{B})_{\omega}$ is not vanishing.
This is not an issue in principle, though care must be taken in order to ensure the largest possible magnitude of the axion source (see Appendix~\ref{sec:Gfactor}).

An SRF production cavity is unable to support EM fields greater than the critical field at which $Q$ severely degrades due to flux penetration.
This sets a fundamental limit on the strength of an SRF axion source which is independent of the cavity $Q$, the input power, or choice of modes.
The limit depends only on the material properties of the chosen superconductor.
For a standard niobium SRF cavity~\cite{Posen:2015amw}, the field limit is
\begin{equation}
(\vec{E} \cdot \vec{B})_\text{srf} \lesssim (0.2~\text{T})^2.
\end{equation}
By comparison, the axion source produced by an RF cavity in a large static field (as in CROWS) is
\begin{align}
(\vec{E} \cdot \vec{B})_\text{rf} \sim (0.1~\text{T})^2\l \frac{P_\text{input}}{100~\text{W}} \r^\frac{1}{2} \l \frac{Q_\text{pc}}{10^5} \r^\frac{1}{2} \l \frac{B_0}{5~\text{T}} \r.
\end{align}
Interestingly an SRF axion source may be similar in magnitude to that of a conventional LSW setup.
The improved reach of our set-up is primarily due to the increase in $Q$ on the detection side, and the decision to employ an SRF or RF cavity for production would depend on more detailed engineering considerations.

\subsection{Gapped toroid conversion region}
\label{sec:toroid}

\begin{figure}
\centering
\includegraphics[width=8cm]{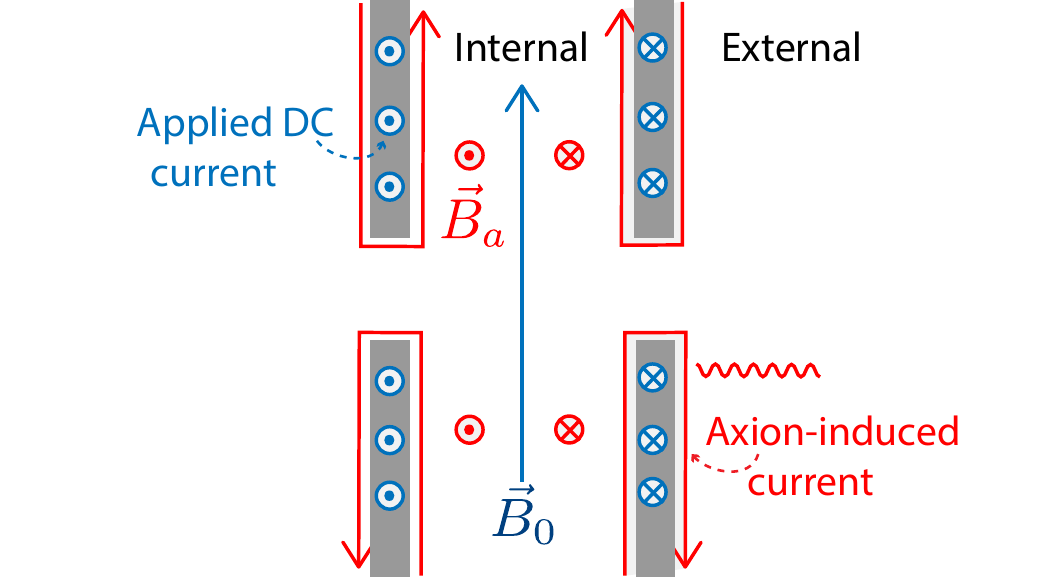}
\caption{Schematic of the gapped toroid as a polarizer, zoomed on to the cross-section of a gap.
The static $\vec{B}_0$ due to applied DC current (blue) remains internal, while the axion-induced $\vec{B}_a$ (red) causes Meissner screening currents (also red) to flow on internal and external surfaces due to the gap. The external currents give rise to detectable fields outside the toroid.
}
\label{fig:polarize}
\end{figure}

An axion interacts with the $\vec{B}_0$ within our conversion region and induces EM fields, described to leading order by effective sources
\begin{align}
\label{eq:Jeff-rhoeff}
\rho_\x{eff} = - g \vec{B}_0 \cdot \vec{\nabla} a,
  ~~~~ \vec{J}_\x{eff} = g \vec{B}_0 \partial_t a
\end{align}
%We do not ignore the spacial gradient of $a$ as the sourced axions may be relativistic.
For a toroidal magnet, the static field is of the form $\vec{B}_0 \sim B_0\l r \r \hat{\phi}$ within the volume of the toroid, and ideally vanishes everywhere outside.
This is the principle advantage of using a toroid, as the SRF cavities can be located in regions of nearly vanishing static field.
However it is essential that the toroid be gapped, for instance due to spaces between wire turns.
A gapped toroid of this sort acts as a polarizer, confining the toroidal static field while permitting the poloidal axion-induced field to propagate outside and be detected, as shown in Fig.~\ref{fig:polarize}.
Indeed, this behavior is same reason that a gapped toroid is being employed in~\cite{Chaudhuri:2014dla, Kahn:2016aff}.

We can understand this as follows: the axion effective current $\vec{J}_\text{eff}$ follows the direction of the static toroidal field $\vec{B}_0$ and sources a poloidal field $\vec{B}_a$.
Both fields vanish in the toroid thickness as Meissner screening currents are set up on the internal surface.
The static field requires poloidal surface currents which are unaware of gaps in the toroid---they do not encounter the gaps as they circulate.
For this reason, the static B-field is effectively contained within the toroid.
Any leakage is due to fringe effects, which are suppressed by the small size of the gap and can be made smaller than the critical SRF threshold.
On the other hand, the axion-induced field will drive toroidal currents which are aware of the gaps.
An internal toroidal current must either collect charge on the edges of the gap or propagate onto the external surface of the toroid, where it sources detectable field.
This field is unsuppressed by the gap size, as long as the gap has a sufficiently small parasitic capacitance (see Sec.~\ref{sec:pickup}).

We now make the approximation that all length scales in the setup (cavity sizes, separations, dimensions of the toroid, etc.)
are comparable and of order $\omega^{-1}$.
We additionally assume that the axion-induced poloidal field $B_a$ is able to escape the toroid without suppression, as though the conducting toroid were not present.
This is valid in the quasistatic limit, as we motivate in Sec.~\ref{sec:screening}.
Combining~\eqref{eq:a-field} and \eqref{eq:Jeff-rhoeff}, we find the axion-induced field in the center of the torus has a magnitude
\begin{align}
\label{eq:Ba-param}
B_a &= \frac{g^2 B_\text{pc}^2 B_0}{\omega^2} \beta \\
&\approx 10^{-26}~\text{T}
\l \frac{g~\GeV}{10^{-11}} \r^2
\l \frac{B_\text{pc}}{0.2~\text{T}} \r^2
\l \frac{B_0}{5~\text{T}} \r
\l \frac{\beta}{0.05} \r, \nonumber
\end{align}
where $B_\x{pc}$ is the field amplitude in the production cavity (note, we have simply taken $E_{\text{pc}} \sim B_{\text{pc}} $ in the above estimate) and $\omega \sim 2\pi \, \text{GHz}$.
$\beta$ is a dimensionless form factor which is a function of the cavity modes, cavity and toroid geometries, spatial variation in $\vec{B}_0$, etc.
The size of $\beta$ is estimated in Appendix~\ref{sec:Gfactor}, and we find in principle that it can be made $\OO(0.1)$ in the limit $m_a \ll \omega$.

\subsection{Screening beyond the quasistatic limit}
\label{sec:screening}

The reasoning presented above for the propagation of axion-induced fields outside the gapped toroid is essentially valid for quasistatic frequencies, $R \lesssim \omega^{-1}$, where $R$ is the characteristic dimension of the toroid.
In the low-frequency limit, the axion-induced magnetic field scales as $B_a \propto (R \omega)$ and so we would try make our toroid as large as possible.
However once $R$ becomes larger than $\omega^{-1}$ the axion-induced fields outside the toroid are suppressed (or \emph{screened}), and thus an optimal design would saturate the quasistatic limit $R \sim \omega^{-1}$.
We discuss this in detail in Appendix~\ref{sec:toymodel}; here we will briefly describe the physical reasons for this result.

Beyond the quasistatic limit, the cross-capacitance of the toroid becomes important: radiation across the center will cause currents and charges on one side of the toroid to affect those on the other side.
Meissner currents flowing along the surface of the toroid are no longer approximately uniform; instead, there will be multiple sections of current flowing in opposite directions, with alternating charge buildups in between.
The resulting Meissner currents and charge distribution is spatially modulated and behaves as a multipolar source.
We thus expect the axion-induced fields outside the toroid to drop-off parametrically as a power-law $B_a \propto (R \omega)^{-n}$, due to destructive interference of out-of-phase source contributions.
We show this behavior and calculate $n>0$ explicitly for a toy model of a thin cylindrical conductor in the high-frequency limit in Appendix~\ref{sec:toymodel}.
Thus, we expect it is safe to saturate $R \sim \omega^{-1}$ without concern that there will be a precipitous (e.g., exponential) drop in the external fields for slightly larger sizes or frequencies.
Likewise, we may treat the approximation of $\OO(1)$ field propagation as accurate even at the boundary of the quasistatic limit.

Note that in our setup the internal toroid signal currents will also have significant spatial modulation beyond the quasistatic limit, but for a very different reason: the source axion field~\eqref{eq:a-field} itself varies on length scales of order $\omega^{-1}$ due to the propagator factor, independent of the choice of modes.
In any case, the multipolar screening described in this section is more general and results from satisfying boundary conditions on the superconducting toroid---this would be present even if the axion field were spatially uniform.

\subsection{Pickup and equivalent transducer circuit}
\label{sec:pickup}

\begin{figure*}
  \centering
  \includegraphics[width=7in]{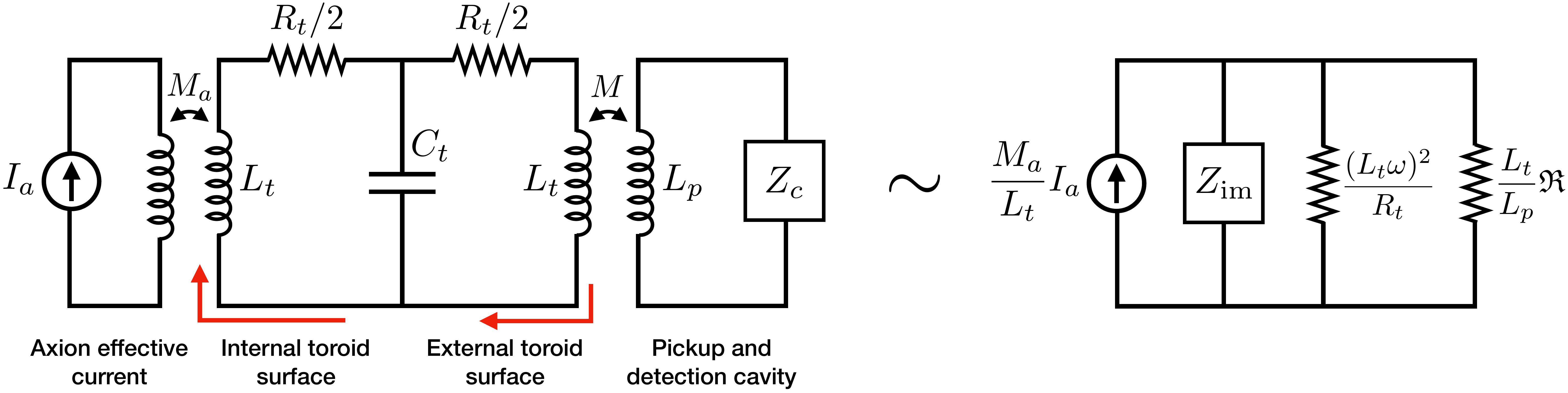}
  \caption{
  (Left) Mapping of our experimental setup onto an effective circuit model.
  This is parameterized by an axion effective current ($I_a$) running through the toroid volume, effective mutual inductance capturing the Meissner effect ($M_a$), toroid inductance ($L_t$), toroid resistance ($R_t$), shunting capacitance ($C_t$), inductive coupling to an outside pickup loop ($L_p$) through a mutual coupling ($M$), and a detection cavity ($Z_c$).
  (Right) Approximate equivalent circuit, for the purposes of computing the maximum signal power. $\mathfrak{R} \propto Q$ is the detection cavity shunt resistance and $Z_\text{im}$ contains all imaginary impedances.
  }
  \label{fig:donut-circuit}
\end{figure*}

To compute the signal strength, it is useful to describe this system with a model circuit, as in the left side of Fig.~\ref{fig:donut-circuit}.
For concreteness we assume the axion-induced EM field is coupled to the detection cavity via a pickup loop located in the central hole of the toroid.
An actual design would likely require a more sophisticated read-out mechanism in order to maintain a large effective $Q$, however this does not alter the optimal signal power.

The model circuit is a straightforward rendering of the signal current flowing on the toroid.
This current flows toroidally, distributed over the inner and outer surfaces of the toroid, as discussed in Sec.~\ref{sec:toroid}. 
We focus on the loop of signal current nearest to the pick-up loop, which flows around the central hole, concentric to the pick-up. 
This current path is represented in the model circuit by the red arrows in Fig.~\ref{fig:donut-circuit}, and it includes segments on both the inner and outer toroid surface.
The magnitude of toroid current is determined by the Meissner boundary conditions. 
It thus receives contributions from the magnetic fields produced by both the axion effect current and any current in the pick-up loop, the latter being a back-reaction which sets the maximal power that may be drawn from the pick-up loop. 

The axion effective current in the volume of the toroid is represented by $I_a$, and its coupling to the inner toroid surface current by an effective mutual inductance $M_a$.
The self inductance of the toroid current path is $L_t \sim R$, with $R$ the toroid radius. 
We choose $M_a \sim R$, which ensures that the current driven in the model circuit due to $I_a$ agrees with that required by the Meissner effect.
The current induced on the inner surface of the toroid can pass to the outer surface, where it couples to a pickup loop of inductance $L_p$ through a mutual inductance $M$ and then feeds into a cavity of impedance $Z_c$.
Alternatively, the current may jump across the gaps between wires and remain on the inner surface: this is captured by the shunting capacitance $C_t$.
As we will show, $\omega L_t \ll (\omega C_t)^{-1}$, and so current always prefers to circulate between the inner and outer surfaces.

The resistances $R_t$ account for the tiny but non-zero surface losses on the toroid.
It is valid to ignore $R_t$ when determining the magnitude of axion-induced fields through~\eqref{eq:Ba-param}.
However, it is important not to ignore it entirely when considering the amplification of signal fields by the SRF cavity.
The detection cavity will be rung up to contain a large current, for which the pickup loop $L_p$ will act as an antenna and excite additional currents on the external surface of the toroid, resulting in additional dissipation via $R_t$.
This back-reaction current is again set by the Meissner boundary conditions, which are reproduced by $M$ and $L_t \sim R$ in the circuit model of Fig.~\ref{fig:donut-circuit}.  
Taking the pick-up loop to have radius $r$ and contain a current $I_p$, it will source a field at the toroid surface $B \sim I_p r^2/R^3$ which requires a Meissner current $I \sim B R \sim I_p r^2/R^2$. 
Since the mutual inductance is of order $M \sim r^2/R$, the required current is indeed $I \sim I_p M/L_t$ that derived from considering our circuit.

We now estimate the relevant model circuit parameters.
The current source $I_a$ represents the total axion effective current threading the toroid and is of order $J_\x{eff} R^2$. 
More precisely, it is the current that gives rise to the outside field $B_a$~\eqref{eq:Ba-param}:
\begin{align}
I_a &\sim B_a R
  \sim \frac{g^2 B_\text{pc}^2 B_0}{\omega^3} \beta \\
 &\approx 10^{-13} \; \x{nA} \; \l \frac{g~\GeV}{10^{-11}}\r^2
  \l \frac{B_\x{pc}}{0.2\;\x{T}} \r^2
  \l \frac{B_0}{5\;\x{T}} \r
  \l \frac{\beta}{0.05} \r \nonumber,
  \label{eq:Ia-estimate}
\end{align}
again with $\omega = 2\pi \, \text{GHz}$ and $m_a \ll \omega$.

Strictly speaking, the two toroid inductances labeled $L_t$ may be different as they inductively couple to different objects.
They are both set by the toroid size, however, so for simplicity we take them both to be
\begin{align}
  L_t \sim R \approx 125~\text{nH} \l\frac{R}{10~\text{cm}}\r.
\end{align}

If the toroid is composed of $N$ turns of wire, then $C_t$ is given by
\begin{align}
  C_t \sim \frac{1}{N} \frac{2 \pi R\cdot d}{g}
\end{align}
where $d$ is the wire diameter and $g = 2 \pi R / N - d$ is the spacing between wires.
For fixed wire diameter, $C_t$ and the fringe fields can be made small by taking a large $N$ and $g \approx d$, which yields:
\begin{align}
    C_t \approx 10^{-2}~\text{pF} \l \frac{d}{\x{mm}} \r.
\end{align}

The use of superconducting wires allows $R_t$ to be as small as few $\text{n}\Omega$ (the minimum RF surface resistance of type II superconductors~\cite{Martinello:2016lrn}), or at worst as large as $\text{m}\Omega$ (the nominal low-temperature resistance of quenched NbTi wires~\cite{Charifoulline:2006sj}).
We expect the resistance will be somewhat larger than $\text{n}\Omega$, as the wires operate in the vortex state and harbor toroidal magnetic flux tubes.
These tubes interact with RF currents in the wires via the so-called Magnus force~\cite{PhysRevLett.70.2158}, and their resulting motion is a significant source of dissipation~\cite{Annett}.
The precise value of $R_t$ will depend on the detailed geometry of the flux tubes and the surface current.
We provide a rough estimate of this resistance, but stress that in what follows we consider the consequences of any $R_t$ within the above bounds.
Since the interaction of RF currents and flux tubes is of the ``Lorentz" form $\vec{J} \times \vec{B}$, the resistance should scale as $\sin\theta$, the angle between the direction of flux tubes and that of the RF current.
In this system, the magnetic field inducing the flux tubes is toroidal but the axion-induced current is poloidal, and so ideally $\theta=0$.
However, the flux tubes will not be perfectly toroidal: static fringe fields provide a deflection of order $\theta \sim B_f/B_0$.
ABRACADABRA has measured the fringe fields outside of their toroidal magnet to be $10^{-6}$ of the primary field~\cite{Oulletconvo}, which we adopt here.
We assume that the deflected component of flux tubes contribute an RF resistance similar to that of trapped flux in SRF cavities, which is on the order $\text{n}\Omega/\text{mG}$~\cite{Martinello:2016lrn}.
Thus we estimate:
\begin{equation}
R_t \approx 100~\text{n}\Omega \l \frac{B_0}{5~\text{T}} \r \l \frac{\theta}{10^{-6}} \r.
\end{equation}

Finally, we choose to model the cavity as a parallel RLC circuit for concreteness, with capacitance $C$, inductance $L$, resistance $\mathfrak{R}$, and thus an impedance:
\begin{align}
Z_c &= \left( \frac{1}{\mathfrak{R}} + \frac{1}{i \omega L} + i \omega C \right)^{-1}.
\label{eq:cavity-impedance}
\end{align}
This cavity has natural resonance frequency $\omega_0^2 = (L C)^{-1}$ and quality factor $Q = \mathfrak{R} / \omega_0 L \gg 1$.
We take $\omega_0 \sim L^{-1} \sim C^{-1} \sim 2\pi \, \text{GHz}$, as set by the physical cavity size.
Note that the effective shunt resistance $\mathfrak{R}$ of this cavity model is very large, proportional to the inverse of the small resistivity of the cavity walls.

All the circuit parameters have so far been estimated by physical considerations, except the pickup loop inductance $L_p$.
This is a free parameter which we tune to optimize the signal, within reasonable limitations as discussed in Sec.~\ref{sec:optimal-signal}.
We assume the mutual inductance $M$ can be made close to optimal, $M \approx \sqrt{L_p L_t}$.
There is also some freedom in choosing the frequency $\omega$ sourced by the production cavity.
Indeed, $\omega$ need not be exactly equal the detection cavity's natural frequency $\omega_0$, although we require that both lie in the GHz range.

\subsection{Optimal signal strength}
\label{sec:optimal-signal}

The signal we are able to extract is given by the power dissipated in the detection cavity.
Here we compute the maximum of this power, varying the pickup inductance and driving frequency.
We use our model circuit for this, and employ the equivalent circuit shown on the right side of Fig.~\ref{fig:donut-circuit}.
This circuit is constructed such that, to lowest order in the small quantities $R_t$ and $C_t$, the power dissipated in the resistor $(L_t / L_p) \mathfrak{R}$ is the same as the power dissipated in the cavity impedance $Z_c$.
Similarly, the power dissipated in the resistor $(L_t \omega)^2 / R_t$ is the same as the total power dissipated in the toroid resistors.

This can be demonstrated by making a series of transformations to sub-circuits of the circuit on the left side of Fig.~\ref{fig:donut-circuit}, each of which preserves the input, output, and dissipated power of the transformed sub-circuit and results in a purely parallel topology.
First, the leftmost transformer can be replaced by a rescaled current source $I_a M_a/ L_t$ and inductor $L_t$.
Recall that $M_a \sim L_t$, so the rescaled current is $\OO \l I_a \r$.
Next the elements between the transformers can be rewritten to lowest order in $R_t$ and $C_t$ as a resistor $(L_t\omega)^2 / R_t$ and capacitor $C_t$.
Finally, the rightmost transformer and cavity impedance can be replaced by an inductor $L_t$ and a rescaled cavity $(L_t / L_p) Z_c$.
The imaginary impedances are gathered into $Z_\text{im}$, which to lowest order in $R_t$ and $C_t$ is:
\begin{align}
  Z_\text{im} &\sim \left( \frac{2}{i\omega L_t} + i\omega C_t + \frac{1}{i\omega \frac{L_t}{L_p}L} + i \omega\frac{L_p}{L_t}C \right)^{-1}.
\end{align}

The system is on resonance when $Z_\text{im}^{-1} = 0$.
To lowest order in $C_t$, this occurs at the frequency
\begin{align}
\omega_\text{res} & \sim  \omega_0\sqrt{1+2\frac{L}{L_p}},
\end{align}
which we will choose to be our driving frequency $\omega$.
On resonance, all current in the equivalent circuit passes through the two resistors.
The power dissipated in the cavity resistor is maximized when these two resistors are equal, which occurs at a pickup loop inductance of
\begin{align}
\label{eq:matched_LP}
   L_\star \; \sim L \; \frac{Q R_t}{L_t \omega_0}.
\end{align}
An inductance $L_p$ that is significantly less than $L$ results in a resonance frequency that is far perturbed from the natural one.
In a realistic experimental implementation, care would need to be taken to ensure that the loaded resonance frequency was not too far perturbed from the detection cavity's natural frequency, lest the quality factor degrade.
As a heuristic implementation of this, we will demand that $\omega \sim \omega_0$ and thus $L_p \gtrsim L$.

We will consider the optimal signal power in two parameter regimes.
First, suppose the cavity is of higher quality than the toroid, $\mathfrak{R} = Q \omega_0 L \gg 1/R_t$.
Impedance matching requires $L_p = L_\star \gg L$, happily yielding a resonance frequency very close to $\omega_0$.
We then draw the \emph{toroid-limited} power
\begin{align}
\label{eq:P_max_DonutLimited}
    P_\x{max} & \sim \frac{1}{8} |I_a|^2 \frac{(L_t \omega)^2}{R_t}.
\end{align}
This is the maximal power that can be extracted from the toroid as long as the driving frequency remains near $\omega_0$.
It thus depends only on the toroid properties and frequency, and notably does not scale with $Q$.

In the second case, suppose that the toroid is of higher quality than the cavity, $\mathfrak{R} = Q \omega_0 L \ll 1/R_t$.
We would hope to again match $L_p$ to $L_\star$, however that would require $L_p \ll L$ and we are thus prevented from impedance matching.
Insisting on $L_p \gtrsim L$, the optimal choice is $L_p \sim L$ for which we draw the \emph{cavity-limited} power
\begin{align}
  \label{eq:P_max_CavityLimited}
P_\text{max} \sim \frac12 |I_a|^2 QL_t\omega_0.
\end{align}

In general, the maximum signal power is the lesser of~\eqref{eq:P_max_DonutLimited} and~\eqref{eq:P_max_CavityLimited}, being limited by resistive losses in the toroid or cavity, respectively:
\begin{align}
\label{eq:combined-maxP}
  P_\text{signal} &\sim |I_a|^2 (\omega L_t)
  \; \x{Min} \left[ \frac{\omega L_t}{R_t}, Q \right]
\end{align}
The relevant toroid parameter to be compared with $Q$ is
\begin{align}
\frac{\omega L_t}{R_t} \sim 10^{10}\l\frac{100~\text{n}\Omega}{R_t}\r.
\end{align}
Thus for $Q \gtrsim 10^{10}$ the toroid impedance may indeed be non-negligible.
The numerical similarity between $Q$ and $\omega L_t/R_t$ reflects the fact that both arise from the small resistivity of superconductors to RF currents.
This also suggests that the experimental details which affect the losses in these systems will be important in determining which of the above regimes is realized.

%%%%%%%%%%%%%%%%%%%%%%%%%%%%%%%%%%%%%%%%%%%%%%%%%%%%%%%%%%%%%%%
\section{Sensitivity to Axion-Photon Coupling}

\subsection{Noise}
\label{sec:noise}

The fundamental sources of noise in this system are thermal and quantum fluctuations of current in the toroid and detection cavity, as well as the intrinsic noise of the device which reads the amplified signal from the cavity.
The thermal and quantum noise can be estimated from the circuit on the right side of Fig.~\ref{fig:donut-circuit}.
The equivalent resistances of both the cavity and toroid will source Johnson currents, behaving as additional parallel current sources.
With $L_p$ tuned as outlined in Sec.~\ref{sec:optimal-signal}, the noise sourced by the effective cavity resistance is always greater than or equal to that sourced by the toroid resistance, so we take a noise source $I_T$:
\begin{align}
\label{eq:thermal-current-noise}
    \left\langle \left| I_T \right|^2 \right\rangle &\sim
    4 T_\x{sys} \frac{1}{\mathfrak{R}} \frac{L_p}{L_t} \; d\nu.
\end{align}
The system temperature $T_\x{sys}$ is the sum of the thermal temperature $T$ and the quantum noise temperature $T_\x{QM} \sim \omega \approx 50~\x{mK}$.
$I_T$ drives fluctuations of the physical magnetic flux $\Phi_T$ inside the detection cavity,
\begin{align}
    \left| \Phi_T \right| =
    \frac{\mathfrak{R}}{\omega_0} \sqrt{\frac{L_t}{L_p}} \left| I_T \right|
\end{align}
resulting in a noise spectrum of cavity flux,
\begin{align}
\label{eq:tsys_noise}
    S_\Phi^{1/2} &\sim \l 4 T_\x{sys} \frac{Q L}{\omega_0}\r^\frac{1}{2} \\
    &\approx \frac{\Phi_0}{\sqrt{\x{Hz}}}
    \l \frac{T_\x{sys}}{0.1~\x{K}} \r^\frac{1}{2}
    \l \frac{Q}{10^{10}} \r^\frac{1}{2} \nonumber
\end{align}
where $\Phi_0$ is the fundamental magnetic flux quantum.

Consider coupling the small signal flux in the cavity to a low-noise read-out device, such as a SQUID magnetometer.
The intrinsic flux noise in such devices is of order $ 10^{-6} \, \Phi_0/\sqrt{\x{Hz}}$~\cite{PhysRevLett.110.147002}, much smaller than the cavity fluctuations~\eqref{eq:tsys_noise}.
We thus take~\eqref{eq:tsys_noise} as the dominant source of noise.

\subsection{Projected sensitivity}
\label{sec:sensitivity}

The noise power extracted from the cavity due to the fluctuations~\eqref{eq:thermal-current-noise} is
\begin{align}
\label{eq:noise-power}
    P_\text{noise} = \mathfrak{R} \frac{L_t}{L_p}
    \left\langle \left| I_T \right|^2 \right\rangle \sim
    4 T_\x{sys} \; d\nu
\end{align}
and the signal-to-noise ratio (SNR) thus
\begin{align}
\label{eq:SNR}
  \text{SNR} \sim \frac{1}{8} |I_a|^2
  \, (\omega L_t) \; \x{Min} \left[ \frac{\omega L_t}{R_t}, Q \right] \frac{t_\text{int}}{T_\text{sys}}
\end{align}
where the relevant bandwidth $d\nu$ is given by the inverse of the total integration time $t_\text{int}$.

One may be concerned that tuning $L_p$ to as outlined in Sec.~\ref{sec:optimal-signal} to maximize the power draw is not truly optimal, as the best measurement will result from maximizing the SNR.
The signal and noise powers extracted from the detection cavity for a general $L_p$, derived from the circuit on the right side of Fig.~\ref{fig:donut-circuit}, are
\begin{align}
\label{eq:general-signal-power}
  P_\text{signal} &\sim \left|I_a\right|^2 \frac{\l L_t \omega\r^2}{R_t}
   \l \frac{L_\star}{L_p} \r \l 1 + \frac{L_\star}{L_p} \r^{-2}, \\
\label{eq:general-noise-power}
  P_\text{noise} &\sim T_\x{sys} \; d\nu \;
  \l 1 + \frac{L_\star}{L_p} \r^{-1}.
\end{align}
The SNR thus nominally increases with decreasing $L_p$, although it saturates to the intrinsic SNR of the toroid at the impedance matched $L_p = L_\star$.
The optimal choice of $L_p$ is thus either $L_\star$ or $L$, the same as that which draws the maximal power~\eqref{eq:combined-maxP}.

\begin{figure}
  \begin{center}
\includegraphics[width=8cm]{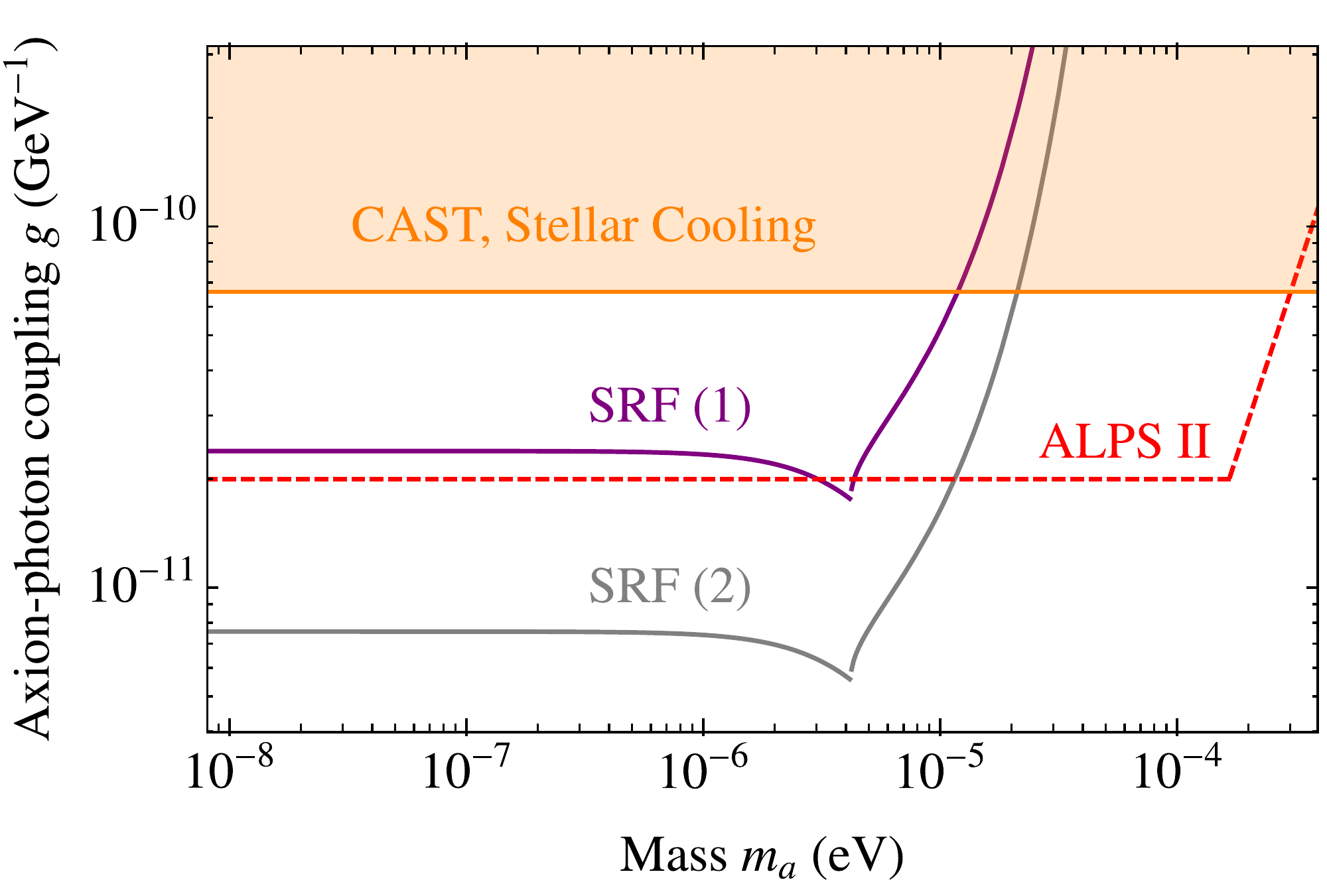}
\end{center}
\caption{Projected sensitivity of proposed SRF LSW setup to axion-photon couplings---see text for the choices of experimental parameters.
Also shown are existing solar axion (CAST)~\cite{Anastassopoulos:2017ftl} and stellar cooling bounds and, for comparison, the future projected reach of the next generation optical experiment ALPS II~\cite{Bahre:2013ywa}.
}
\label{fig:sensitivity}
\end{figure}

Demanding $\text{SNR} > 5$, the estimated reach at low axion masses $m_a \ll \omega$ is given by:
\begin{align}
\label{eq:greach}
g > 2 \cdot 10^{-11}~\GeV^{-1} \cdot
\l \frac{\omega/2\pi}{\x{GHz}}\r
\l \frac{B_0}{5~\x{T}} \r^{-\frac{1}{2}}&  \nonumber \\
\l \frac{B_\x{pc}}{0.2~\x{T}} \r^{-1}
\l \frac{\beta}{0.05} \r^{-\frac{1}{2}}
\l \frac{L_t}{125~\x{nH}} \r^{-\frac{1}{2}}& \nonumber \\
\l \frac{R_t}{100~\x{n}\Omega} \r^{\frac{1}{4}}
\l \frac{t_\x{int}}{\x{year}} \r^{-\frac{1}{4}}
\l \frac{T_\x{sys}}{0.1~\x{K}} \r^{\frac{1}{4}}&.
\end{align}
This is independent of the detection cavity quality factor if it is sufficiently large ($Q \geq 10^{10}$ for these parameters).
The full sensitivity is show in Fig.~\ref{fig:sensitivity} using:
\begin{align*}
&\frac{\omega}{2\pi} = \text{GHz}, ~~ B_\text{pc} = 0.2~\text{T}, ~~B_0 = 5~\text{T} \\
&L_t = 125~\text{nH}, ~~t_\text{int} = 1~\text{year}, ~~ T_\text{sys} = 0.1~\text{K},
\end{align*}
and considering two cases of cavity and toroid losses:
\begin{align*}
&(1)~R_t = 100~\x{n}\Omega~~\x{and}~~ Q \geq 10^{10} ~~ \\
&(2)~R_t = \x{n}\Omega~~\x{and}~~ Q \geq 10^{12}.
\end{align*}
We have used a form factor of $\beta = 0.05$, assuming $m_a \ll \omega$ (see Appendix~\ref{sec:Gfactor}).
The estimated sensitivity of our SRF axion design is capable of surpassing current astrophysical limits, and is comparable to the expected reach of the next generation optical experiment, ALPS II~\cite{Bahre:2013ywa}.

%%%%%%%%%%%%%%%%%%%%%%%%%%%%%%%%%%%%%%%%%%%%%%%%%%%%%%%%%%%%%%%
\section{Discussion}
\label{sec:discussion}

We have proposed a novel design for an LSW axion search leveraging SRF cavity technology and employing a region of isolated, static magnetic field.
Our particular realization uses a gapped toroid, similar to that of~\cite{Chaudhuri:2014dla, Kahn:2016aff}, to contain a static field while allowing the propagation of axion-induced signal fields.
It would be interesting to consider other possible geometries for the conversion region, though the gapped toroid illustrates the necessary features.
Our focus in this work is understanding the fundamental factors which set the sensitivity of such an experiment, namely the possible screening of the signal fields beyond the quasistatic limit and back-reaction from the non-negligible toroid impedance.
We calculate the optimal signal strength, and for reasonable toroid parameters and SRF quality factors we find a sensitivity to axion-photon couplings in excess of astrophysical limits and comparable to complementary optical experiments.
Notably, the optimal sensitivity is in fact independent of both production and detection cavity $Q$ factors in the limit of large $Q$, and is instead determined by the properties of the conversion region.

We conclude with a few comments on experimental feasibility that have not yet been addressed.
We have modeled the coupling of the detection cavity and axion-induced signal fields with an inductive pickup, yet a naive implementation of such a coupling would likely compromise the detection $Q$ due to losses in the pickup wire.
It is critical to explore coupling mechanisms that will not degrade $Q$, which is complicated by the fact that the toroid operates on the extreme of the quasistatic regime and thus requires microwave engineering.
There are other sources of noise not considered here which must be understood and managed in a practical implementation, such as stray external fields which require shielding and additional losses due to non-superconducting support materials used in the system.
Finally, perhaps the biggest engineering challenge here is the necessity of frequency-matching the two SRF cavities to within $1/Q \lesssim 10^{-10}$.
This demands a precise resonance monitoring and feedback mechanism to counter frequency drifts, and is a major hurdle for any photon regeneration experiment utilizing high-$Q$ cavities, such as~\cite{Graham:2014sha}.

\section*{Acknowledgments}

We would like to thank Karl van Bibber, Saptarshi Chaudhuri, Jeff Dror, Peter Graham, Kent Irwin, Jonathan Ouellet, Ben Ponedel, Sam Posen, Nick Rodd, Ben Safdi, and Jesse Thaler for valuable discussions. S.R.~was supported in part by the NSF under grants PHY-1638509 , the Simons Foundation Award 378243 and the Heising-Simons Foundation grants  2015-038 and 2018-0765. 

%%%%%%%%%%%%%%%%%%%%%%%%%%%%%%%%%%%%%%%%%%%%%%%%%%%%%%%%%%%%%%%
\begin{appendices}

\section{Estimate of the axion-induced fields}
\label{sec:Gfactor}

In this section we estimate the magnitude of the axion-induced fields, assuming a simple geometry for the production cavity and toroidal conversion region.
From the expressions for the axion source field~\eqref{eq:a-field} and effective current~\eqref{eq:Jeff-rhoeff}, the axion-induced magnetic field at a detection point $\vec{r}$ is generally of the form:
\begin{align}
\label{eq:Ba-full}
  &\vec{B}_a(\vec{r}) = \frac{i \omega g^2}{\l4 \pi\r^2} \, e^{i \omega t}
   \int_\x{pc} d^3 y \, \int_\x{cr} \, d^3 x \;\; \\
  &\left\{\vec{\lambda} \times \vec{B}_0\l\vec{x}\r \nonumber
  \left[ \frac{1}{\lambda^3} + \frac{i \omega}{\lambda^2} \right]
  \frac{e^{-i \omega \lambda} \, e^{i k_a |\vec{x} - \vec{y}|}}
  {|\vec{x} - \vec{y}|} \l \vec{E}\cdot\vec{B} \r_{\omega}\right\}.
   \end{align}
Here the integration $\vec{y}$ is taken over the volume of the production cavity, $\vec{x}$ is over the volume of the conversion region, and $\vec{\lambda} = (\vec{r} - \vec{x})$ is the separation vector between points in the toroid and a detection point $\vec{r}$.
The time-dependent $J_\text{eff}$ has been evaluated at the retarded time $t_r = t - \lambda$.
\eqref{eq:Ba-full} also uses the approximation that the axion-induced fields fully propagate outside of the toroid, as expected for quasistatic frequencies.
$B_a$ lies in the poloidal direction and has an amplitude:
\begin{align}
\label{eq:Ba-beta}
\hat{z} \cdot \vec{B}_a = \frac{g^2 B_\text{pc}^2 B_0}{\omega^2} \beta (\vec{r}),
\end{align}
where the dimensionless form factor $\beta$ contains information about the choice of cavity modes, etc.

First we specify the dimensions involved.
Consider a circular cylindrical cavity (``pill-box") of radius $a$ and height $h$.
The resonant frequencies are
\begin{align}
\omega^\text{TM}_{npq} &= \sqrt{\l \frac{x_{np}}{a} \r^2+ \l\frac{q \pi}{h}\r^2} \\
\omega^\text{TE}_{npq} &= \sqrt{\l \frac{x'_{np}}{a} \r^2+ \l\frac{q \pi}{h}\r^2}, \nonumber
\end{align}
for $\text{TM}_{npq}$ and $\text{TE}_{npq}$ modes respectively, where $x_{np}$ and $x'_{np}$ are the $p$th roots of the $n$th order Bessel function $J_n(x)$ and its derivative $J_n'(x)$~\cite{Hill}.
Setting $a = h/2 = 10~\text{cm}$ ensures resonant frequencies of order $\approx \text{GHz}$ for low-lying modes, typical of SRF cavities.

Next consider a toroid of inner radius $R$ and rectangular cross section of height and width $R$.
We take the cylindrical cavity to be aligned axially with the toroid, with a minimal separation distance of $(h+R)/2$.
Though this should be gapped toroid, we can approximate the static field contained inside the toroidal volume as
\begin{equation}
\vec{B}_0 (r) = B_0 \l \frac{R}{r} \r \hat{\phi},
\end{equation}
for $r \in [R, 2R]$ where $r$ is the cylindrical radial distance from the center.
If we require the toroid size saturates the quasistatic limit $R \omega \sim 1$, an economical choice for the dimension is simply $R = a$.

We now consider the axion source in this setup.
The source axion field is greatest when $\vec{E} \cdot \vec{B}$ is maximal and coherent throughout the production cavity volume.
Since we have assumed a cylindrical cavity with no external field, it is necessary to drive multiple modes to ensure a non-vanishing $(\vec{E} \cdot \vec{B})_{\omega}$.
The choice of these modes is not obvious and requires care even in this simple setup.

To demonstrate an ill-advised choice consider the $\text{TM}_{010}$ and $\text{TM}_{111}$ modes which results in $(\vec{E} \cdot \vec{B})_{\omega} \propto \sin \l\pi z/h\r \sin(\phi)$.
Note that the integral of $(\vec{E} \cdot \vec{B})_{\omega}$ vanishes over $z \in [-h/2,h/2]$ of the production cavity.
This $z$-dependence is in fact a general feature of any cylindrical cavity modes chosen, but it is not detrimental as we are operating in the near-field regime.
Rather, $\text{TM}_{010}$ and $\text{TM}_{111}$ represents a poor choice of modes because of the $\phi$ dependence---the sourced axion field will be purely harmonic in the azimuthal angle, and thus would integrate over the toroid to give a highly suppressed signal field near the center.
This cancellation is essentially a consequence of the symmetry and alignment of the cylindrical setup and is easily avoidable.
One potential solution is to place the production cavity in a position off the axial axis.
Another is to modify the toroid wiring so $\vec{B}_0$ also varies with the azimuthal angle while still being effectively confined.
One can also select cavity modes such that $(\vec{E} \cdot \vec{B})_{\omega}$ is not purely harmonic in $\phi$: the lowest-lying combination of cylindrical modes which yields this angular behavior is the $\text{TM}_{111}$ and $\text{TE}_{111}$ modes.

In any case, we can estimate a reasonable upper limit to $\beta$ in \eqref{eq:Ba-beta} by postulating a perfectly \emph{uniform} $\vec{E} \cdot \vec{B}$ throughout the production cavity volume.
Taking this optimal axion source, we numerically find that $\beta$ is roughly constant for points in the center of the toroid:
\begin{equation}
\beta_\text{optimal} \approx 7 \cdot 10^{-2},~~~z=0 ~\text{and}~r \leq R.
\end{equation}
Here we have also taken the limit in which the mass is negligible, $m_a \ll \omega$.
At masses $m_a \gtrsim \omega$, there is the usual exponential drop-off from producing off-shell axions.
If we instead use a perhaps more realistic axion source by driving the $\text{TM}_{111}$ and $\text{TE}_{111}$ combination, we numerically find that $\beta_\text{realistic} \approx 4 \cdot 10^{-4}$, again roughly constant near the center of the toroid.

In summary, we expect the form factor $\beta$ can in principle be made $\OO(0.1)$ in any suitably engineered designs.
As discussed, it is important to determine a suitable geometry and choice of modes to be driven in the SRF production cavity, as a poor choice could lead to a significant suppression of the signal fields.

\section{A toy model for screening}
\label{sec:toymodel}

We ultimately rely on the quasistatic approximation in assuming the axion-induced fields propagate $\OO(1)$ out of the gapped toroid, similar to~\cite{Kahn:2016aff}.
This limits the size of the toroid to be less than or of order the inverse frequency of the axion.
However it is important to understand the \emph{degree} to which the fields outside the toroid are suppressed at larger frequencies or larger toroid size.
This is a complicated boundary-value problem and a full study would require a detailed numerical computation which is outside the scope of this work.
We will demonstrate here the power-law nature of this suppression.

To gain some intuition, consider an electromagnetic field of frequency $\omega$ impinging on a perfect conducting sheet.
If the conductor is infinitely large, then the incoming field is reflected and vanishes on the far side of the conductor (i.e., metals are shiny).
An analogous behavior holds for fields sourced inside of a region bounded by a closed conducting surface---the field is exactly screened outside (i.e., phones do not work in elevators).
The common feature is that the conductors lack a boundary.
We thus expect incident fields to be suppressed, but not exactly screened, outside of a large yet finite conductor with a definite boundary.
This will occur when the conductor size $H$ is much larger than the wavelength $\omega^{-1}$.

Now suppose the conductor is small relative to $\omega^{-1}$.
This is just the quasistatic limit, so we may asses the conductor's response by considering its response to a static field.
In this familiar situation, the field will induce charges and currents on the surface of the conductor in order to screen the bulk.
It is clear that the boundaries play an important role in this limit.
For example, a conducting block in a static electric field will develop a screening charge density on the boundary, which modifies the net external field but does not result in a parametrically small external field.
For $\omega H \ll 1$ we therefore expect the field on the far side of the conductor to only differ from the incident field by $\OO(1)$ factors.

We study here a toy model of electromagnetic fields incident on a finite cylindrical conductor.
The parametric effects of screening can be sensibly extracted in the high-frequency limit, and we find the magnitude of external, detected fields are only power-law suppressed compared to the internal fields.
The physical mechanism underlying this suppression, as summarized in Sec.~\ref{sec:screening}, is expected to hold generically in varied geometries.

\begin{figure}
\centering
\includegraphics[width=8cm]{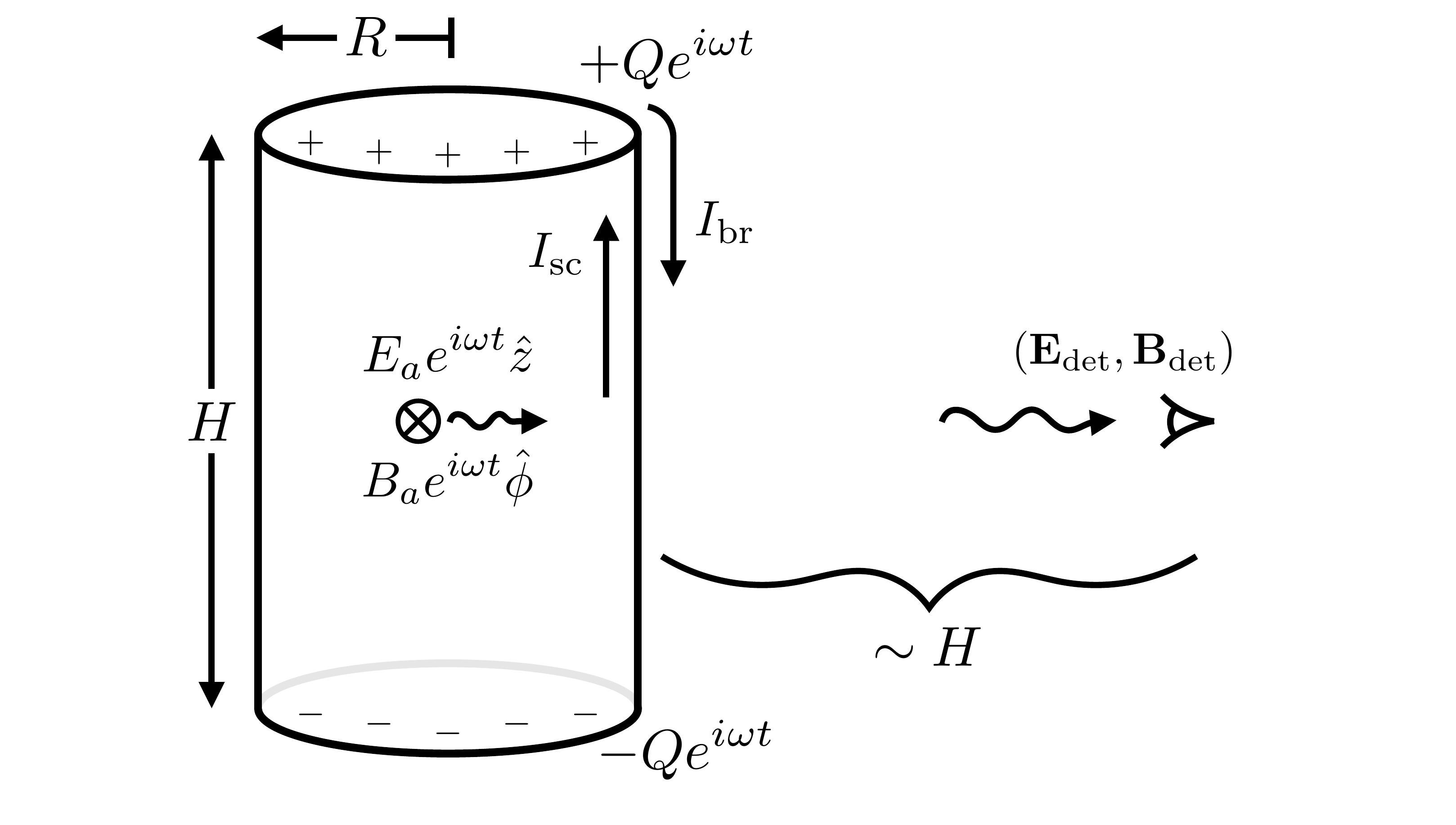}
\caption{Screening of EM fields $(\vec{E}_a, \vec{B}_a)$ sourced inside a finite cylindrical wall.
In the high-frequency limit, the conductor response results in charge buildup $Q$ on the edges and a configuration of screening currents $I_\text{sc}$, $I_\text{br}$ on the inner and outer surfaces.
These in turn determine the fields $(\vec{E}_\text{det}, \vec{B}_\text{det})$ detected outside the cylinder.
}
\label{fig:toy}
\end{figure}

Consider a perfectly conducting cylinder of height $H$ and radius $R$.
More precisely, take this to be a tube of negligible thickness separating an inner and outer cylindrical wall.
Suppose there is an EM field $(\vec{E}_a, \vec{B}_a) = (E_a\hat{z}, B_a\hat{\phi})$, sourced by an infinite line of current $I_a e^{i \omega t} \hat{z}$ ``in the throat" of the cylinder.
This is labelled suggestively in analogy to fields sourced by the axion interaction with a static magnetic field, although for simplicity we assume a spatially uniform $I_a$.
We specifically examine the limit of a thin cylinder and take $R \sim \omega^{-1} \ll H$, which is of course well beyond the quasistatic approximation.
Here the fields radiated by $I_a$ are cylindrical plane waves, with approximate magnitudes:
\begin{equation}
E_a(r) \sim  B_a(r) \sim \omega I_a e^{i \omega t} \l \frac{1}{\omega r} \r^{1/2}, ~~ R \lesssim r \lesssim H.
\label{eq:nonmodulationhighfreq}
\end{equation}
These source fields will be compared to the detected fields $(\vec{E}_\text{det}, \vec{B}_\text{det})$ at a point $ r \sim H$ outside the cylinder.
This is depicted in Fig.~\ref{fig:toy}.
From here on, we restrict our attention to the behavior of fields in the region $R \lesssim r \lesssim H$, extending from the cylindrical surface to the detection point.
We will also ignore any contributions to the fields due to the source wire $I_a$ ``sticking out'' the ends,
since this finite cylinder is intended to resemble an ``unwrapped" version of our gapped toroid.

To determine the detected fields, the conductor response is paramount.
Boundary conditions dictate that the electric and magnetic fields vanish in the thickness of the cylinder, and the $z$-component of the electric field vanishes on the surface.
Importantly, for a finite cylinder the inner and outer surfaces are connected, so that the current established on the inner wall is communicated in some form to the outer wall.
This communication, and the detected fields that result, can be estimated by approximately satisfying boundary conditions as follows:

Firstly, $I_a$ drives a screening current $I_\text{sc}$ on the inner walls in order to cancel the source fields.
By continuity, there is then necessarily a charge buildup $\pm Q e^{i \omega t}$ at the top and bottom edges of the cylinder, respectively.
We will not attempt to explicitly satisfy boundary conditions near these edges (which involves complicated fringe effects).
Instead, we will consider the effects of this charge on the rest of the cylinder at locations far from the edges---that is, $H/2-|z| \gg R$, where $z=0$ corresponds to the vertical center of the cylinder.
Here the oscillating rings of charge at $z = \pm H/2$ appear as points and produce an electric field on the cylinder surface:
\begin{align}
&\hat{z} \cdot \vec{E}_Q(R,z) \sim Q e^{i \omega t} \nonumber \\
 &\left( \frac{e^{-i \omega z_+} (1+i \omega z_+)}{z_+^2} + \frac{e^{- i \omega  z_-} (1+i \omega z_-)}{z_-^2} \right),
\label{eq:edge}
\end{align}
where $z_\pm = H/2\, \pm\, z$.
Up to a phase, this is approximately
\begin{align}
\hat{z} \cdot \vec{E}_Q(R,z) \sim \frac{\omega Q e^{i \omega t}}{H/2 - |z|}\cos (\omega z).
\label{eq:EQapprox}
\end{align}
To continue satisfying boundary conditions, this field must now be canceled.
Therefore, a ``back-reaction'' current $I_\text{br}$ must be set up on the cylinder walls, chosen to cancel $\vec{E}_Q$.
Numerically, we find that a current of the form $I_\text{br}(z) \sim I_\text{br} \cos(\omega z)$ sources electric fields with the necessary sinusoidal behavior:
\begin{align}
\hat{z}\cdot \vec{E}_\text{br}(R,z) & \sim \frac{I_\text{br} e^{i \omega t}}{H/2 - |z|} \cos(\omega z).
\label{eq:currentEapproxOnSurface}
\end{align}
We can ensure that this back-reaction does not also violate the previously satisfied boundary conditions by taking $I_\text{br}$ to flow in the same same direction on both inner and outer walls.
It is notable that near the center of the cylinder,~\eqref{eq:currentEapproxOnSurface} vanishes as the height increases $H \to \infty$.
Such a scaling can be understood by considering a $ \cos(\omega z)$ current on the surface of an \emph{infinitely} tall cylinder.
In that case, the $z$-component of the electric field exactly vanishes as there is a cancellation between the field sourced by the current and the field sourced by stripes of charge which are present due to charge continuity.
This cancellation is weaker near the edges of a finite cylinder, leading to larger $\hat{z} \cdot \vec{E}_\text{br}$ there.

The above charges/currents must be self-consistent.
The initial screening current $I_\text{sc}$ on the inner wall, charge buildup $Q$ on the edges, and back-reaction currents $I_\text{br}$ on both walls here are related by charge continuity:
\begin{equation}
\frac{dQ}{dt} = i \omega Q \sim I_\text{sc} - 2I_\text{br},
\label{eq:cont}
\end{equation}
where the factor of $2$ accounts for the fact that $I_\text{br}$ flows in the same direction on both walls.
Since the cylinder is tall and thin, we can invoke the infinite-cylinder solution to approximately cancel the source fields, and thus we take $I_\text{sc} \sim I_a$ on the inner surface.
Comparing~\eqref{eq:EQapprox} and~\eqref{eq:currentEapproxOnSurface}, to cancel the fields produced by the charge buildup requires back-reaction currents of order $I_\text{br} \sim \omega Q$.
These currents, taken together, then approximately satisfy boundary conditions everywhere away from the edges.
Now further demanding the constraint of continuity~\eqref{eq:cont}, we find the charge buildup should be $Q \sim I_a / \omega$, and therefore the back-reaction currents are of order $I_\text{br} \sim I_a\cos (\omega z)$.
The $z$-component of the back-reaction field is parametrically smaller than the source field~\eqref{eq:nonmodulationhighfreq} on the surface:
\begin{align}
\frac{\hat{z} \cdot \vec{E}_\text{br} (R,z)}{E_a (R,z)} \sim (\omega H)^{-1} \ll 1.
\end{align}
This is consistent with our use of the infinite-cylinder solution for the inner screening current $I_\text{sc} \sim I_a$.

To summarize, we have found there are additional currents $I_\text{br} \sim I_a\cos (\omega z)$ on the inner and outer cylinder surfaces, arising from the need to satisfy boundary conditions in the presence of charge build-up.
These are inevitably of the same order as the source current, but with a crucial spatial modulation.
Based on these currents, we estimate the detected fields at a point $r \lesssim H$ (and near $z \sim 0$) outside the cylinder:
\begin{align}
\vec{E}_\text{det}(r,z) &\sim I_a  \cos(\omega z) e^{i \omega t} \l \frac{1}{r} \hat{r} + \frac{1}{H+r} \hat{z}\r, \nonumber \\
\vec{B}_\text{det}(r,z) &\sim \frac{I_a}{r} \cos(\omega z) e^{i \omega t} \hat{\phi}.
\label{eq:currentEapprox}
\end{align}
The fields radiated by the oscillating charges on the cylinder edges are of this same magnitude.

The charge buildup and back-reaction currents thus propagate fields outside the cylinder.
Importantly, the magnitudes of these fields~\eqref{eq:currentEapprox} fall off as $r^{-1}$, faster than the $r^{-1/2}$ behavior of the source fields~\eqref{eq:nonmodulationhighfreq} that would be seen if the conductor were not present.
Comparing these, we see the magnitude of the external, detected field is power-law suppressed:
\begin{align}
B_\text{det}/B_a \sim (\omega H)^{-1/2} \ll 1.
\end{align}
This is fundamentally because the fields radiated by a modulated, multipolar current decay more rapidly than the fields from a spatially uniform current.

Lastly, we briefly comment on the low-frequency behavior of our toy model.
The spatially modulated current distribution we had found on the surface is a consequence of modulated fields from non-negligible charge buildup---this feature, however, is only present at sufficiently high frequencies.
In the opposite limit $R, H \ll \omega^{-1}$, the analogous secondary fields from charge buildup are uniform across the cylinder surface and drive an unmodulated current that results in equal and opposite currents flowing on the inner and outer surfaces.
This is the familiar quasistatic result in which no screening occurs and a uniform current loop is established, as in the operation of a cryogenic current comparator~\cite{Grohmann:1974}.

\end{appendices}

%%%%%%%%%%%%%%%%%%%%%%%%%%%%%%%%%%%%%%%%%%%%%%%%%%%%%%%%%%%%%%%

\bibliography{axion}
\end{document}